\documentclass[a4paper,11pt]{article}

\pdfoutput=1
\usepackage{jcappub}
\usepackage[export]{adjustbox}
\definecolor{linkblue}{rgb}{0,0,0.8}
\definecolor{linkgreen}{rgb}{0,0.5,0}
\hypersetup{linkcolor=linkblue, citecolor=linkgreen, urlcolor=linkblue}

\usepackage{lmodern}
\usepackage{bm}
\usepackage{esint}
\usepackage{verbatim}
\usepackage[us]{datetime}
\usepackage{booktabs}
\usepackage{enumerate}
\usepackage{graphics}
\usepackage{float}
\usepackage{xcolor}
\usepackage[normalem]{ulem}
\graphicspath{{./figs/}}

\begin{document}
\title{\boldmath One-parameter dynamical dark-energy from the generalized Chaplygin gas}

\author{Rodrigo von Marttens,}
\author{Dinorah Barbosa,}
\author{Jailson Alcaniz}
\affiliation{Observat\'orio Nacional, 20921-400, Rio de Janeiro, RJ, Brasil}

\emailAdd{rodrigovonmarttens@gmail.com}
\emailAdd{dinorahteixeira@on.br}
\emailAdd{alcaniz@on.br}

\abstract{The fact that Einstein’s equations connect the space-time geometry to the total matter content of the cosmic substratum, but not to individual contributions of the matter species, can be translated into a degeneracy in the cosmological dark sector. Such degeneracy makes it impossible to distinguish cases where dark energy (DE) interacts with dark matter (DM) from a dynamical non-interacting scenario using observational data based only on time or distance measurements. In this paper, based on the non-adiabatic generalized Chaplygin gas (gCg) model, we derive and study some cosmological consequences of a varying one-parameter dynamical DE parameterization, which does not allow phantom crossing. 
We perform a parameter selection using the most recent public available data, such as the data from Planck 2018, eBOSS DR16, Pantheon and KiDS-1000. We find that current observations  provide strong constraints on the model parameters, leading to values very close to the $\Lambda$CDM cosmology, at the same time that the well-known $\sigma_8$ tension is reduced from $\sim 3\sigma$ to $\sim 1\sigma$ level.}


\maketitle

%%%%%%%%%%%%%%%%%%%%%%%%%
%%%%%%%%%%%%%%%%%%%%%%%%%
\section{Introduction}

Since the discovery of cosmic acceleration from  Type Ia Supernovae (SNe) observations~\cite{perlmutter:1998,riess:1998obs}, one of the main goals of Cosmology has been to uncover the physical nature of such phenomenon~\cite{Durrer:2007re}. In the context of general relativistic models, the current phase of accelerated expansion is due to an exotic negative pressure contribution, credited to a dark energy component. In the standard ${\Lambda}$CDM model, such a component takes the form of a cosmological constant $\Lambda$, described as a perfect fluid with equation of state parameter $w=-1$, and associated to the vacuum energy. However, as well known, there is a huge discrepancy  between the theoretical prediction for the vacuum energy from quantum field theory and its value obtained from cosmological observations. This outstanding disagreement is the main theoretical aspect that ${\Lambda}$CDM fails to explain~\cite{weinberg1989cosmological,Martin:2012bt,Padmanabhan:2002ji}. Likewise, the fact that the contribution of the dark components to the current energetic budget of the Universe reached the same order of magnitude in recent times, is also a theoretical challenge to the standard cosmological model. This intriguing synchronism is commonly referred to as the Cosmic Coincidence Problem~\cite{Velten:2014nra}.

From the observational viewpoint, it is well known that the ${\Lambda}$CDM model provides a successful description of the structure and evolution of the universe. However, the current discrepancy for the Hubble constant $H_0$ value at $\simeq 5 \sigma$ between Cosmic Microwave Background (CMB) measurements and low-$z$ SNe data is particularly challenging (see e.g. \cite{Riess:2022mme,DiValentino:2021izs}). Another ongoing issue concerns a $\simeq 2\sigma$ difference in the $\Omega_m -\sigma_8$ plane that can be inferred by comparing CMB and cosmic shear data \cite{Hildebrandt:2018yau}, where $\Omega_m$ is the matter density parameter and $\sigma_8$ is the matter fluctuation amplitude on scales of $8h^{-1}$ Mpc. 

In this context, testing the limits of the $\Lambda$CDM model~\cite{Andrade:2021njl}, as well as looking for alternative descriptions in order to solve, or at least alleviate, the aforementioned issues~\cite{DiValentino:2021izs} has been of great interest. One of these possibilities concerns on assuming a non-minimal coupling between the energy components of the cosmological dark sector~\cite{Costa:2009wv,vomMarttens:2016tdr,vonMarttens:2018iav,Benetti:2019lxu}. In fact, there is no physical motivation that prevents such coupling, which in turn introduces a new physical mechanism that affects the dynamics of the Universe, both at the background and perturbative levels. These are constantly invoked as an alternative to address the issues faced by the $\Lambda$CDM model~\cite{DiValentino:2020zio,DiValentino:2020vvd}. In general, the explicit form of the interaction is phenomenologically introduced through an \textit{ansatz} of a source four-vector that affects the energy and momentum balance of the dark components, even though the total energy is conserved. 

On the other hand, for a given theory of gravity (i.e., for a given Einstein tensor $G_{\mu\nu}$), we are able to find the total value of the energy-momentum tensor of the cosmic substratum. While we can obtain the energy-momentum tensor for baryons and radiations through direct observations, the same is not true for the dark energy sector whose physical nature is yet to be determined. Since there is an large number of possibilities to describe the dark sector satisfying the field equations, this creates a degeneracy, the so-called \textit{dark degeneracy}. In particular, for interactive scenarios, the dark degeneracy implies that certain interactive models could mimic non-interactive models and vice-versa. As previously shown in \cite{vonMarttens:2019ixw}, it is possible to take advantage of this fact and write a more generalized interaction, ruled by the DM and DE ratio and seek for signatures that only appear in the interactive approach. Here, we focus on the inverse way as originally performed in \cite{vonMarttens:2019ixw}, that is, to obtain the dynamical models for a given interaction. In this work, we study the case of a one-parameter dark energy from the well-known generalized Chaplygin gas (gCg) ~\cite{Bento:2002ps,Alcaniz:2002yt,Cunha:2003vg,Gorini:2007ta,Piattella:2009da,Fabris:2008hy}.

This paper is organized as follows: In Section~\ref{sec:dark_degeneracy} we describe the background framework of the dark degeneracy, making explicit the equations that relate the interacting approach to the (non-interacting) dynamical DE approach. Section~\ref{sec:gcg} applies the obtained equations for the gCg model, which results in the background and perturbative descriptions. Section~\ref{sec:stat_an} presents the observational data used in our analysis as well as the main constraints on the dynamical model parameters. The main conclusions of the paper are discussed in Section~\ref{sec:conclusions} 

%%%%%%%%%%%%%%%%%%%%%%%%%
%%%%%%%%%%%%%%%%%%%%%%%%%
\section{Dark degeneracy}
\label{sec:dark_degeneracy}

%%%%%%%%%%%%%%%%%%%%%%%%%
%%%%%%%%%%%%%%%%%%%%%%%%%
\subsection{Background description of the dark sector}

The cosmological background means the scales where the cosmological principle holds, i.e., the scales where the matter is distributed homogeneously and isotropically. In this regime, the spacetime can be described by the FLRW metric,
\begin{equation} \label{flrw}
ds^{2}=dt^{2}-a^{2}\left(t\right)\left[dr^{2}+r^{2}\left(d\theta^{2}+\sin^{2}\theta d\phi\right)\right]\,,
\end{equation}
and the dynamics of the Universe is described by the Friedmann equation, 
\begin{equation} \label{friedmann}
3H^{2}=8\pi G\rho\,.
\end{equation}
In Eq.~\eqref{friedmann}, $\rho$ is the energy density of the total cosmic substratum, which accounts for all the matter components that compose the Universe. For the material content, we assume that the Universe is composed of radiation, baryons, cold dark matter (CDM), and dark energy (DE). In the background, each one of these components is described as a perfect fluid with equation of state (EoS) $p_{i}=w_{i}\rho_{i}$, where the EoS parameter $w_{i}$ is not necessarily a constant. The radiation component will be denoted by the subindex $r$ and is characterized by $w_{r}=1/3$ whereas the baryonic and CDM components will be denoted by the subindexes $b$ and $c$, respectively, and are characterized by $w_{b}=0$ and $w_{c}=0$. Since we will consider an interaction only between the dark components, baryons and CDM must be considered separately~\cite{vomMarttens:2014ftc}. Lastly, the DE component will be denoted by the subindex $x$, and we consider a time-dependent EoS parameter $w_{x}\left(a\right)$.

\subsection{Unified dark sector}

In the standard cosmological scenario, all components are considered independent, i.e., there is no physical process able to provide an energy or momentum exchange between the components. As mentione above, 
here we relax this assumption, but not wholly: we assume a possible interaction between the dark components with baryons and radiation conserving separately, so that ${\rho}_{b} \propto a^{-3}$ and ${\rho}_{r} \propto a^{-4}$.

It is always possible to effectively describe the background of the dark sector as a unified fluid, even if it is physically described as separated (interacting or not) components. The energy density and pressure of the dark fluid are given by
\begin{equation} \label{rhop}
\rho_{d}=\rho_{c}+\rho_{x}\quad,\quad p_{d}=p_{x}=w_{x}\left(a\right)\rho_{x}\,,
\end{equation}
where the subscript $d$ denotes the unified dark fluid. 

Using the energy density described in Eq.~\eqref{rhop}, we can write the unified dark energy density in terms of the ratio $r=\rho_{c}/\rho_{x}$ and the DE energy density,
\begin{equation}
\rho_{d}=\big[1+r\left(a\right)\big]\rho_{x}\,.
\end{equation}
Since the dark pressure is only due to the DE component, we can write an effective EoS for the unified dark fluid,
\begin{equation} \label{EoS}
p_{d}=w_{d}\left(a\right)\rho_{d}\qquad {\rm with}\qquad w_{d}\left(a\right)=\frac{w_{x}\left(a\right)}{1+r\left(a\right)}\,.
\end{equation}
Note that the unified energy density is the sum of the dark components' energy densities, so that the dark sector as a whole must be conserved. This  means that the unified dark fluid must satisfy the background conservation equation,
\begin{equation} \label{dark}
\dot{\rho}_{d}+3H\big[1+w_{d}\left(a\right)\big]\rho_{d}=0\,.
\end{equation}
For a general EoS parameter $w_{d}\left(a\right)$, the conservation equation has a well-known solution,
\begin{equation} \label{rhod}
\rho_{d}\left(a\right)=\rho_{d0}\exp\left[-3\int\frac{1+w_{d}\left(a^{\prime}\right)}{a^{\prime}}da^{\prime}\right]\,,
\end{equation}
which has been explored to assess the possibility of a dark sector interaction in a model-independent way~\cite{vonMarttens:2018bvz,vonMarttens:2020apn}.

Since the background dynamics is described by the Friedmann equation \eqref{friedmann}, no individual component is relevant alone, but the sum of all energy densities. Thus, at the background level, all the dark sector information is contained in $\rho_{d}$ or, according to equation \eqref{rhod}, in the dark EoS parameter $w_{d}\left(a\right)$. 

As can be seen in equation \eqref{EoS}, the unified dark EoS parameter depends directly on the functions $w_{x}\left(a\right)$ and $r\left(a\right)$. Whereas the first one is related to the dynamical nature of the DE component, the second one, as it is discussed in the Appendix~\ref{ssec:intDE}, can be associated to an interaction in the dark sector via equation \eqref{dr}. However, using equation \eqref{EoS}, it is clear that different combinations of $w_{x}\left(a\right)$ and $r\left(a\right)$ can give us exactly the same $w_{d}\left(a\right)$ and, consequently, the same Hubble rate. Therefore, no observation based only on measurements of time and distances, e.g., cosmic chronometers and SNe Ia, can distinguish models with the same $w_{d}\left(a\right)$.

It is important to emphasize that this degeneracy does not imply that different descriptions with the same $w_{d}\left(a\right)$ but different $w_{x}\left(a\right)$ and $r\left(a\right)$ are completely equivalent, since the dark components are different separately: whereas the CDM evolves with $a^{-3}$ in the dynamical DE parameterization approach, the interaction affects CDM evolution in the second case.

%%%%%%%%%%%%%%%%%%%%%%%%%
%%%%%%%%%%%%%%%%%%%%%%%%%
\subsection{Equivalence between dynamical DE and interacting dark sector}

As aforementioned, the dark degeneracy is a well-known concept in which different approaches for describing the dark sector can be degenerated from the point of view of the observations~\cite{Wasserman:2002gb,Rubano:2002sx,Kunz:2007rk,Aviles:2011ak,Carneiro:2014uua}. Let us consider the following two approaches:
\begin{itemize}
	\item \textbf{Dynamical DE: }As well as the $\Lambda$CDM model, here we consider that the dark components are independent, i.e., they do not interact with each other. However, instead of $w_x=-1$, DE component is described by a time-dependent EoS parameter $w\left(a\right)$. Several parameterization for $w\left(a\right)$ have already been studied in the cosmological context (e.g., the CPL~\cite{Chevallier:2000qy,Linder:2002et},  Barboza-Alcaniz~\cite{Barboza:2008rh} and Wetterich parameterizations~\cite{Wetterich:2004pv}). From now on, all quantities related to the dynamical approach will be denoted with a bar. A brief description of the dynamics of the Universe in this approach is presented in Appendix~\ref{ssec:dynDE}.
	\item \textbf{Interacting DE: }As well as the $\Lambda$CDM model, here we consider that the DE component is described by $w_x=-1$. However, instead of independent components, we consider that CDM and DE interact. More precisely, the interaction is introduced via a source term in the covariant derivative of the dark components. At the background level, the interaction is described by a time-dependent function $r\left(a\right)$, as the ratio between CDM and DE energy densities. Several interacting models have already been studied in the cosmological context~(see e.g. \cite{Cid:2018ugy} and references therein). From now on, all quantities related to the interacting approach will be written with a tilde. A brief description of the dynamics of the Universe in this approach is presented in Appendix~\ref{ssec:intDE}.
\end{itemize}
%To make this relation explicit at the background level, we shall formally obtain the conditions that must be satisfied in order to get the same Hubble rate for both approaches discussed in Secs.~\ref{ssec:dynDE} and~\ref{ssec:intDE}. We consider that radiation and baryons evolve in both cases as $a^{-3}$ and $a^{-4}$, respectively, so that all differences between the dynamical and interacting approaches come from the dark sector's dynamics. 

It is possible to establish a map between these two approaches. From Eq.~\eqref{rhod}, it is straightforward to conclude that the degeneracy is verified for all models with the same EoS parameter for the unified dark fluid $w_{d}\left(a\right)$. Considering only the dynamical and interacting approaches, this requirement can be explicitly formulated as $\bar{w}_{d}\left(a\right)=\tilde{w}_{d}\left(a\right)$, which, using Eqs.~\eqref{wd1} and~\eqref{wd2}, reduces to
\begin{equation} \label{degeneracy}
\tilde{r}\left(a\right)=-\dfrac{1+\bar{r}_{0}a^{-3}\exp\left[3\int\dfrac{1+\bar{w}_{x}\left(a^{\prime}\right)}{a^{\prime}}da^{\prime}\right]}{\bar{w}_{x}\left(a\right)}-1\;.
\end{equation}
The above expression is the key equation in the context of dark degeneracy because it establishes an explicit mapping between dynamical and interacting approaches. As an example, let us start from a model with dynamical DE, characterized by $\bar{w}_{x}\left(a\right)$. According to Eq.~\eqref{degeneracy}, it is always possible to find an interacting description, characterized by $\tilde{r}\left(a\right)$, which gives exactly the same background expansion dynamics. In this case, since now the ratio $\tilde{r}\left(a\right)$ is known, the interacting function $\tilde{f}\left(\tilde{r}\right)$ can be obtained using Eq.~\eqref{dr}
\begin{equation} \label{ftilde}
\tilde{f}\left(\tilde{r}\right)=-\dfrac{\tilde{r}^{\prime}a}{3\tilde{r}}-1\,,
\end{equation}
where prime denotes the derivative with respect to scale factor. This procedure has been used, for example, in Ref.~\cite{vonMarttens:2019ixw} to build analogous interacting models for three well-known DE parameterizations: $w$CDM, CPL~\cite{Chevallier:2000qy,Linder:2002et}, and Barboza-Alcaniz~\cite{Barboza:2008rh}. 

A remarkable feature of this mapping between dynamical and interacting approaches is that the solutions for the dark component's energy densities satisfy an algebraic relation. Using Eqs.~\eqref{rhop} and~\eqref{EoS}, the degeneracy condition can be written as
\begin{equation} \label{degen}
\bar{w}_{x}\left(a\right)\dfrac{\bar{\rho}_{x}}{\bar{\rho}_{c}+\bar{\rho}_{x}}=-\dfrac{\tilde{\rho}_{x}}{\tilde{\rho}_{c}+\tilde{\rho}_{x}}\,.
\end{equation}
Since the total unified dark energy density must be equal in both approaches, $\rho_{d}=\bar{\rho}_{c}+\bar{\rho}_{x}=\tilde{\rho}_{c}+\tilde{\rho}_{x}$\footnote{No bar or tilde is used in the definition of $\rho_{d}$ because this quantity is identical in both approaches by construction.}, the denominators in Eq.~\eqref{degen} are identical, and then, it leads to the following relations
\begin{eqnarray}
\tilde{\rho}_{c}&=&\bar{\rho}_{c}+\bar{\rho}_{x}\left[1+\bar{w}_{x}\left(a\right)\right]\,, \label{rhoc2} \\
\tilde{\rho}_{x}&=&-\bar{w}_{x}\left(a\right)\bar{\rho}_{x}\,. \label{rhox2}
\end{eqnarray}

Eqs.~\eqref{rhoc2} and~\eqref{rhox2} indicate how the solutions of the energy densities in the interacting (analogous) model can be found if the energy densities and the EoS parameter are known in the dynamical approach. In this work, we desire to take the opposite direction, i.e., from an interacting model, we want to find its dynamical counterpart. In that sense, Eqs.~\eqref{rhoc2} and~\eqref{rhox2} can be inverted to obtain
\begin{eqnarray}
\bar{\rho}_{x}&=&\tilde{\rho}_{c}-\bar{\rho}_{c}+\tilde{\rho}_{x}\,, \label{rhoc3} \\
\bar{w}_{x}&=&-\frac{\tilde{\rho}_{x}}{\tilde{\rho}_{c}-\bar{\rho}_{c}+\tilde{\rho}_{x}}\,. \label{rhox3}
\end{eqnarray}
Note that, when we start from the interacting approach, the two unknown quantities are $\bar{\rho}_{x}$ and $\bar{w}_{x}$, but CDM energy density evolves with $a^{-3}$. In this case, Eq.~\eqref{degeneracy} is an integral equation, which means that there is one extra degree of freedom that must be fixed. 
%
%At this point, before imposing the additional constraint, it is convenient to redefine the interacting parameter as
%\begin{equation} \label{xiwx}
%    \tilde{\gamma}=1+\tilde{w}_{0}\,.
%\end{equation}
%Since $\tilde{\gamma}$ is assumed to be constant, $\tilde{w}_{0}$ is also constant. A more general case, where $\tilde{\gamma}$ is time-dependent can be formulated, but it is out of the scope of this work. This variable change is particularly convenient because in both approaches the $\Lambda$CDM limit is recovered when $\bar{w}_{x}=\tilde{w}_{0}=-1$.
%
%In principle, there are several options to eliminate the extra degree of freedom mentioned earlier. 
Here we choose the following condition
\begin{equation} \label{wx0}
    \bar{w}_{x}\left(a=1\right)=\bar{w}_{0}\,.
\end{equation}
It worth mentioning that, even though degenerated models have identical expansion rates, the predicted values for the current value of CDM and DE density parameters are different ($\bar{\Omega}_{c0}\neq\tilde{\Omega}_{c0}$ and $\bar{\Omega}_{x0}\neq\tilde{\Omega}_{x0}$). This difference can be easily understood taking Eqs.~\eqref{rhoc3} and~\eqref{rhox3} at $z=0$. 

%%%%%%%%%%%%%%%%%%%%%%%%%
%%%%%%%%%%%%%%%%%%%%%%%%%
\section{Decomposed generalized Chaplygin gas model}
\label{sec:gcg}

The generalized Chaplygin gas (gCg) model is characterized by the EoS ~\cite{Bento:2002ps,Alcaniz:2002yt}
\begin{equation}
    p_{ch}=-\frac{A}{\rho_{ch}^{\alpha}}\,,
\end{equation}
where $A$ is a strictly positive constant and $\alpha$ is a constant free parameter that must satisfy the condition $\alpha>1$. It has been applied to cosmology as an attempt to unify the dark sector~\cite{Kamenshchik:2001cp,Fabris:2001tm,Bilic:2001cg,Dev:2002qa} whose solution for the energy density is given by 
\begin{equation} \label{eosgcg}
    \rho_{ch}=\left[A+\frac{B}{a^{3\left(1+\alpha\right)}}\right]^{\frac{1}{1+\alpha}}\,,
\end{equation}
where $B$ is a constant of integration. It is worth mentioning that cosmological models based on the gCg have been widely studied in literature~\cite{Cunha:2003vg,Gorini:2007ta,Piattella:2009da,Fabris:2008hy}, although its adiabatic version presents instabilities in the matter power spectrum, as shown in~\cite{Sandvik:2002jz}. On the other hand, when non-adiabatic perturbations are considered, the instabilities are removed~\cite{Reis:2003mw,HipolitoRicaldi:2009je,HipolitoRicaldi:2010mf}, making them a viable alternative to the dark sector. 

One way to introduce non-adiabaticities to the gCg model is to decompose it into two segments, corresponding to the usual two dark components: CDM and DE. In this concern, a viable proposal to perform such decomposition consists in dividing the gCg into interacting dark components, which is called {decomposed generalized Chaplygin gas}~\cite{Zhang:2004gc,Wands:2012vg}. In this interacting scenario, the decomposed gCg model has already undergone several observational tests~\cite{Li:2013bya,Wang:2013qy,Wang:2014xca,Marttens:2017njo}. We refer the reader to Refs.~\cite{Benetti:2019lxu,Benetti:2021div} for an updated analysis on the parameter selection for the decomposed gCg with the most recent available data.

\subsection{(Interacting) Decomposed gCg}

First, let us consider the interacting decomposition of the gCg. In what follows, we consider that the energy density and pressure of the gCg are divided into CDM and DE contributions:
\begin{equation}
    \rho_{ch}=\tilde{\rho}_{c}+\tilde{\rho}_{x}\quad{\rm and}\quad p_{ch}=\tilde{p}_{x}\,.
\end{equation}
We also impose that the DE component is described by a vacuum EoS, i.e., $\tilde{w}_{x}\left(a\right)=\tilde{p}_{x}/\tilde{\rho}_{x}=-1$. As indicated by the tildes, the components are supposed to interact with each other, which means that the background energy conservation is given by Eqs.~\eqref{cdmenergy} and~\eqref{deenergy}. As already developed in the Refs.~\cite{Zhang:2004gc,Wands:2012vg}, the decomposed gCg is described by the following source function\footnote{Since we are interested in degenerated models, we omit the tilde notation in the Hubble parameter.}
\begin{equation} \label{Qgcg}
\tilde{Q}=3H\left(1+\tilde{w}_{0}\right)\dfrac{\tilde{\rho}_{c}\, \tilde{\rho}_{x}}{\tilde{\rho}_{c}+\tilde{\rho}_{x}}\,,
\end{equation}
which leads to the following solutions for the background energy densities,
\begin{eqnarray}
\tilde{\rho}_{c}&=&\frac{8\pi G}{3H_{0}^{2}}\tilde{\Omega}_{c0}a^{3\tilde{w}_{0}}\left(\dfrac{\tilde{\Omega}_{c0}a^{3\tilde{w}_{0}}+\tilde{\Omega}_{x0}}{\tilde{\Omega}_{c0}+\tilde{\Omega}_{x0}}\right)^{-1-\frac{1}{\tilde{w}_{0}}}\,, \label{rhoctildegcg} \\
\tilde{\rho}_{x}&=&\frac{8\pi G}{3H_{0}^{2}}\tilde{\Omega}_{x0}\left(\dfrac{\tilde{\Omega}_{c0}a^{3\tilde{w}_{0}}+\tilde{\Omega}_{x0}}{\tilde{\Omega}_{c0}+\tilde{\Omega}_{x0}}\right)^{-1-\frac{1}{\tilde{w}_{0}}}\,. \label{rhoxtildegcg}
\end{eqnarray}

The Eqs.~\eqref{rhoctildegcg} and \eqref{rhoxtildegcg} fully determine the background dynamics of the Universe in the interacting decomposed gCg scenario.

\subsection{(Dynamical) Decomposed gCg}
\label{ssec:dyncGc}

From now on, we will call the dynamical analogous model to the decomposed gCg as $\bar{w}$gCg model. In practice, this can be understood as a different decomposition of the gCg, but imposing that dark components do not interact with each other and allowing DE component to be dynamical. 

In order to obtain the background solutions in the dynamical approach, we combine Eqs.~\eqref{rhoc3} and~\eqref{rhox3}, where the CDM energy density in the dynamical approach is given by Eq.~\eqref{rhoc1}, and the interacting solutions are given by Eqs.~\eqref{rhoctildegcg} and~\eqref{rhoxtildegcg}, with the condition~\eqref{wx0}. This procedure leads us to the following solutions for the DE energy density and DE EoS parameter:
\begin{eqnarray} 
\bar{\rho}_{x}&=&\frac{8\pi G}{3H_{0}^{2}}a^{-3}\left[\left(\tilde{\Omega}_{c0}+\tilde{\Omega}_{x0}\right)^{\frac{1}{\tilde{w}_0}+1}\left(\tilde{\Omega}_{x0}a^{-3\tilde{w}_0}+\tilde{\Omega}_{c0}\right)^{-\frac{1}{\tilde{w}_0}}-\frac{\tilde{w}_0 \left(\tilde{\Omega}_{c0}+\tilde{\Omega}_{x0}\right)+\tilde{\Omega}_{x0}}{\tilde{w}_0}\right]\,, \label{rhoxdyn} \\
\bar{w}_{x}&=& - \frac{\tilde{w}_0\tilde{\Omega}_{x0}(\tilde{\Omega}_{c0} + \tilde{\Omega}_{x0})^{1 + \frac{1}{\tilde{w}_0}}}{(a^{3\tilde{w}_0}\tilde{\Omega}_{c0} + \tilde{\Omega}_{x0})\Big(\tilde{w}_0(\tilde{\Omega}_{c0} + \tilde{\Omega}_{x0})^{1 + \frac{1}{\tilde{w}_0}} - (\tilde{\Omega}_{c0} + a^{-3\tilde{w}_0}\tilde{\Omega}_{x0})^\frac{1}{\tilde{w}_0}(\tilde{\Omega}_{x0} + \tilde{w}_0(\tilde{\Omega}_{c0} + \tilde{\Omega}_{x0})) \Big)}\,. \nonumber \\
&& \label{wxdyn}
\end{eqnarray}
The Eqs.~\eqref{rhoxdyn} and~\eqref{wxdyn} fully determine the background dynamics of the Universe in the dynamical decomposed gCg scenario. 

There are some important aspects of the dynamics of the $\bar{w}$gCg model that can be analyzed from those equations. First, from~\eqref{wxdyn}, we can conclude that, at $z=0$, the DE EoS results in $\tilde{w}_0$. This means that the today's value of the DE EoS parameter $\bar{w}_{x}\left(z=0\right)=\bar{w}_{0}$ is identical to the interaction parameter $\tilde{w}_0$. Thus, from now on we also omit the bar/tilde in the $w_0$ parameter. In essence, this parameter reflects the deviation from $\Lambda$CDM model, independently of the approach. Second, the Eqs.~\eqref{rhoxdyn} and~\eqref{wxdyn} are written in terms of the ``interacting'' parameters (with tildes). {{In reality, this is a mathematical convenience because in the dynamical scenario, only the ``dynamical'' parameters (with bars) have physical meaning}}. For example, the density parameters for CDM and DE components in the dynamical approach are given by
\begin{equation} \label{Omegabar}
    \bar{\Omega}_{c0}=\frac{w_{0} \tilde{\Omega}_{c0}+w_{0} \tilde{\Omega}_{x0}+\tilde{\Omega}_{x0}}{w_{0}}\quad{\rm and}\quad\bar{\Omega}_{x0}=\frac{\tilde{\Omega}_{x0}}{w_0}\;.
\end{equation}
It is possible to verify that Eq.~\eqref{wxdyn} only produces physical solutions for $-1<w_{0}<0$, and in all cases the EoS parameters is restricted to $w_{0}\leq\bar{w}_{x}\left(z\right)\leq 0$. In practice, this means that the $\bar{w}$gCg model does not admit phantom DE solutions and also does not allow that the EoS parameter crosses the value $-1$. In this respect, our approach delivers an one-parameter dynamical EoS for the DE component that naturally avoids the phantom regime.

\begin{figure*}[h]
\centering
\includegraphics[width=0.49\columnwidth]{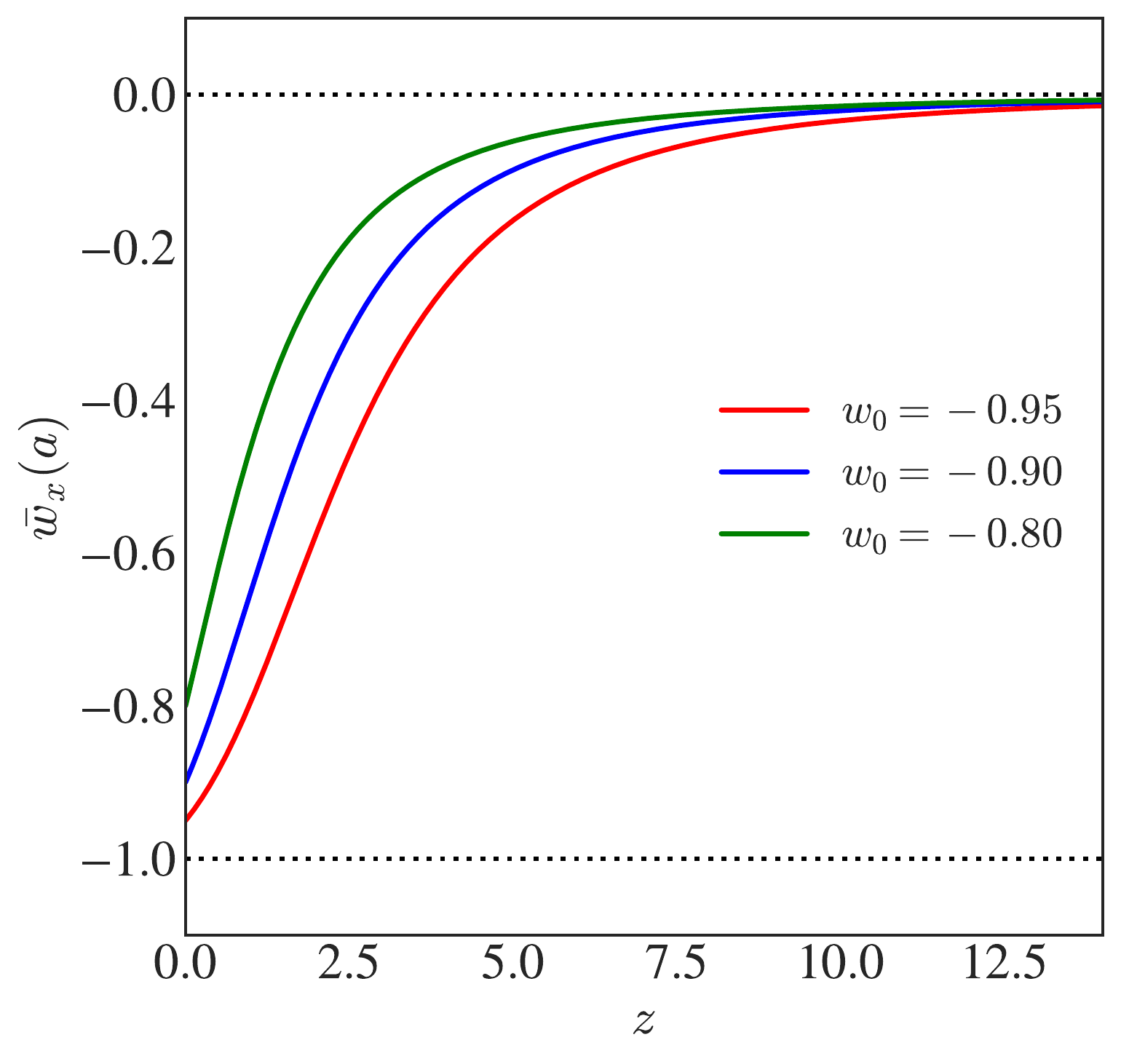}
\includegraphics[width=0.48\columnwidth]{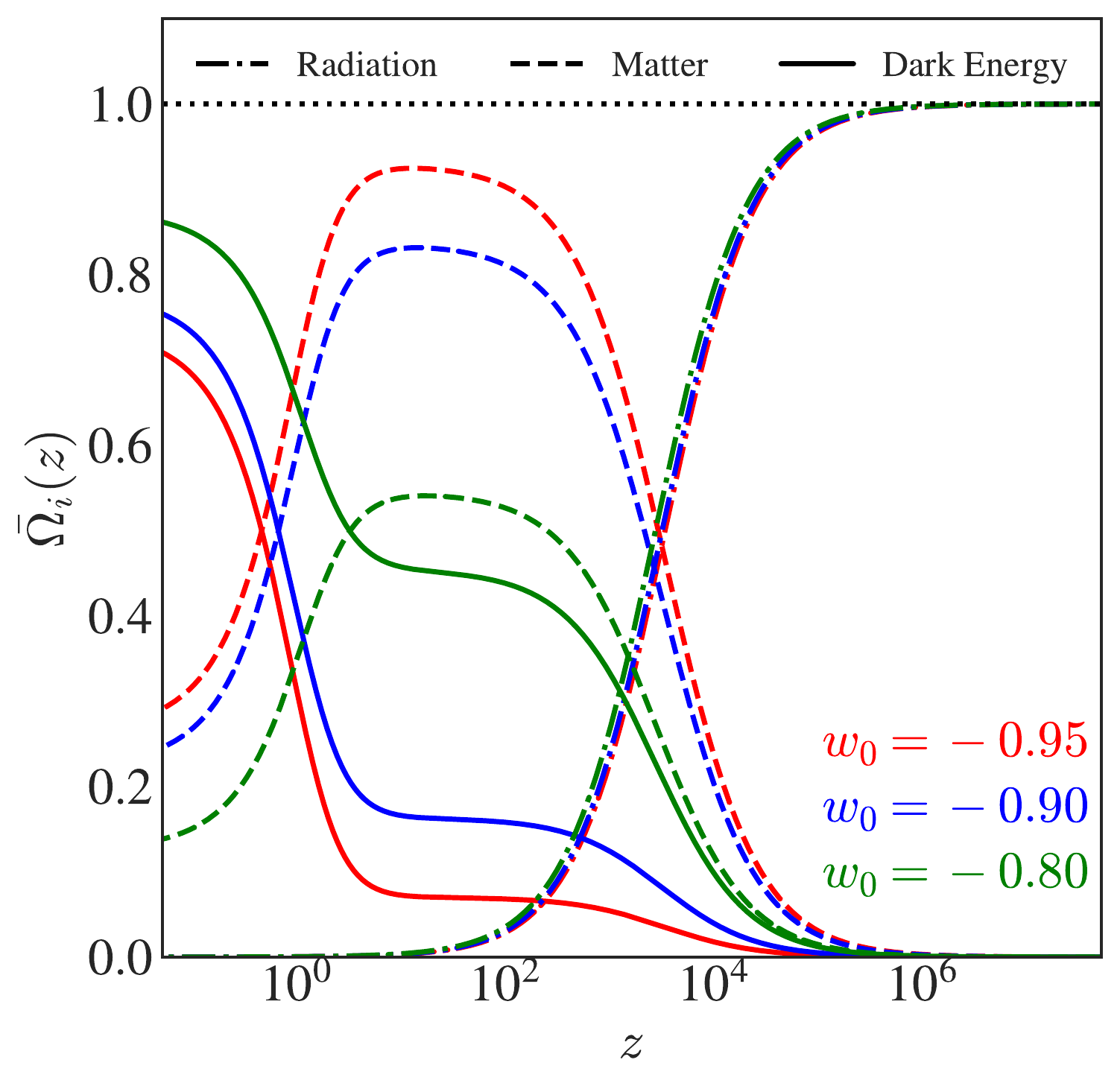}
\caption{Background solutions for the $\bar{w}$gCg model for different values of $w_0$. Both plots consider $\tilde{\Omega}_{c0}=0.25$, and the value of $\bar{\Omega}_{c0}$ can be obtained from Eq.~\eqref{Omegabar}. \textbf{Left panel: }EoS parameter in the dynamical approach for different values of $w_0$. In each case, the dark energy component behaves as matter in the early universe and transits to a negative $\bar{w}_x$ around $z = 0.1$. Due to physical bounds, $\bar{w}_x$ never reaches values below $-1$. \textbf{Right panel: }Density parameter for all matter species. The radiation component is denoted by the dashed-dotted line: the matter component, which contains CDM and baryons, is denoted by the dashed line; and the DE component is denoted by the solid line.}
\label{fig.wx}
\end{figure*}

In order to compute the background dynamics of the $\bar{w}$gCg, we make use of a suitable modified version of the Boltzmann solver CLASS~\cite{Blas:2011cosmic} with the equations developed in Sec.~\ref{ssec:dyncGc}. In Fig.~\ref{fig.wx}, we show the background solutions for the $\bar{w}$gCg model considering some specific cases varying the parameter $w_0$. Whereas in the left panel of Fig.~\ref{fig.wx} we show the time evolution of the EoS parameter of the DE component, in the right panel of Fig.~\ref{fig.wx} we show the time evolution of the density parameter for all species. From this figure, it is possible to conclude that the DE component has a pressureless phase in the early Universe ($\bar{w}_{x}\approx 0$), shows a transient evolution to a negative EoS starting from $z=10$, and reaches the value $\tilde{w}_{0}$ today, as imposed by the condition~\eqref{wx0}. This pressureless phase creates a ``step'' in the evolution of the DE component. As will be discussed latter, this behavior will be important for our perturbative analysis of the models. A remarkable feature of this model is that if we look to the matter component only between $z=10^{0}$ and $z=10^{3}$, we conclude that matter density parameter never reaches the totality (i.e., $\bar{\Omega}_{m}=1$). At first glance, this could indicate that there is no matter-dominated epoch in the $\bar{w}$gCg model. However, in this period, the DE component is pressureless, which means that, at the background level, it behaves like matter in this period, making a late-time transition to a negative pressure fluid.

\subsection{Perturbations}

Since our analysis aims at testing the $\bar{w}$gCg model with the most recent data, which includes the CMB data from Planck (TTTEEE+lensing), we must compute the first-order perturbative equations. In particular, we need to assess the consequences of the peculiar behavior of the DE component, showed in Fig.~\ref{fig.wx} in the perturbative dynamics.

Beyond the (perturbative) Einstein's equations, the relevant equations to understand the evolution of the perturbations of a given matter species individually are the first order energy-momentum conservation equations. Considering linear perturbation around FLRW metric in Newtonian gauge, where the metric fluctuations are described by the potentials $\phi$ and $\psi$, the conservation equations for a matter species denoted by a sub-index $i$ are
\begin{eqnarray}
    \delta_{i}^{\prime}&=&-\left(1+w_{i}\right)\left(\theta_{i}-3\phi^{\prime}\right)-3\mathcal{H}\left(c_{s,i}^{2}-w_{i}\right)\delta_{i}\,, \label{pertdelta} \\
    \theta_{i}^{\prime}&=&-\mathcal{H}\left(1-3w_{i}\right)\theta_{i}-\frac{w_{i}^{\prime}}{1+w_{i}}\theta_{i}+\frac{c_{s,i}^{2}}{1+w_{i}}k^{2}\delta_{i}+k^{2}\psi\,, \label{perttheta}
\end{eqnarray}
where $\delta_{i}\equiv\delta\rho_{i}/\rho_{i}$ is the density contrast and $\theta\equiv ik^{j}v_{j}$ is the divergence of the total spatial 3-velocity of the matter species $i$. Furthermore, an important quantity in Eqs.~\eqref{pertdelta} and~\eqref{perttheta} is the rest-frame sound speed $c_{s,i}^{2}$. Note that Eqs.~\eqref{pertdelta} and~\eqref{perttheta} are written in terms of the conformal time, so that prime denotes derivatives w.r.t. conformal time and $\mathcal{H}\equiv a^{\prime}/a$. 

The perturbations for the $\bar{w}$gCg were also calculated by CLASS. More precisely, for the (pressureless) CDM component, we adopt, as usual, $w_c=0$ and $c_{s,c}^{2}=0$. On the other hand, for the DE component we use Eq.~\eqref{wxdyn} for the EoS parameter, and, motivated by the scalar field sound speed, we fix $c_{s,x}^{2}=1$. Such a high value for the DE sound speed prevents DE clustering, due to the pressure action. 

\begin{figure}[h]
\centering
\includegraphics[width=0.49\columnwidth]{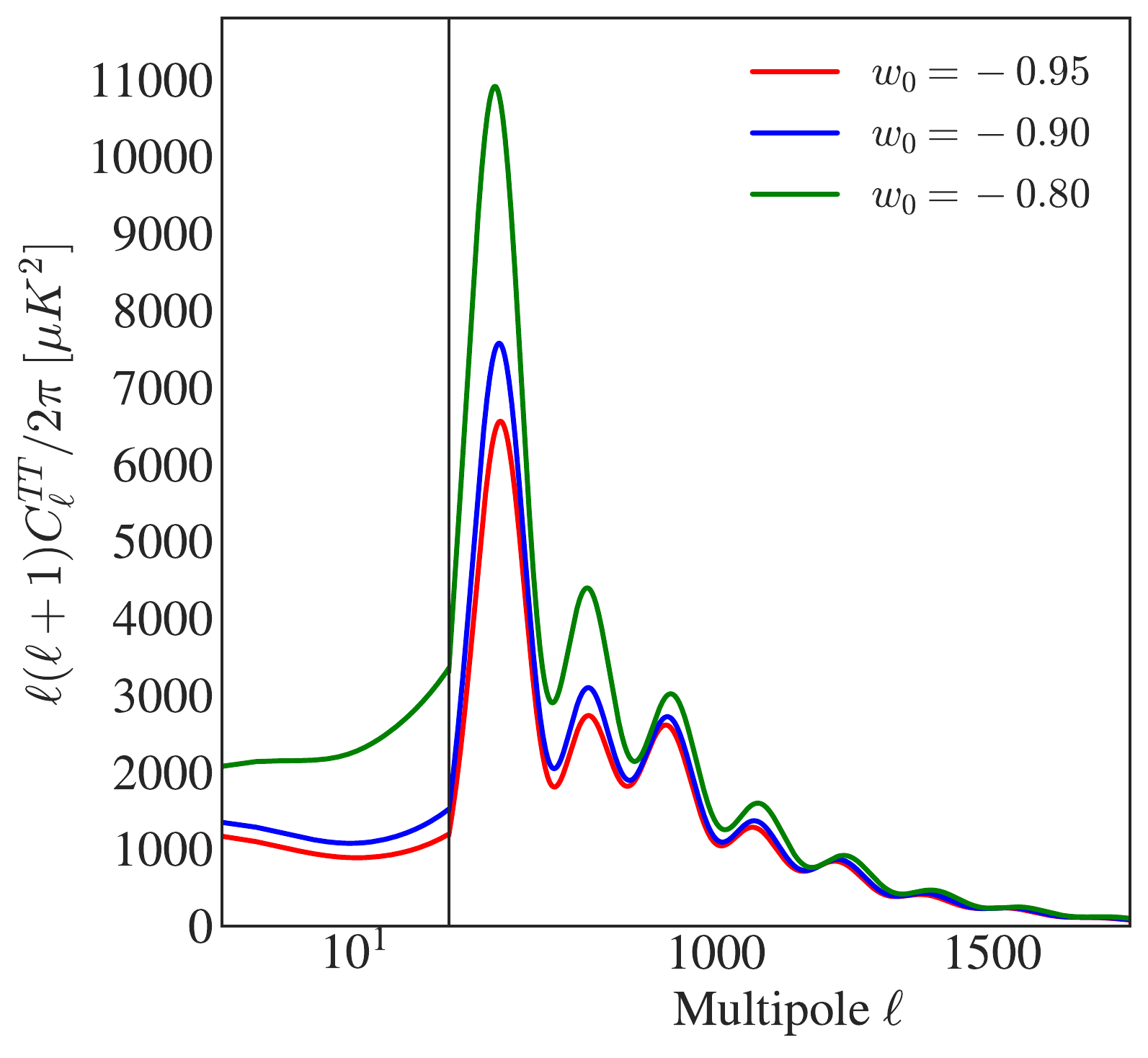}
\includegraphics[width=0.48\columnwidth]{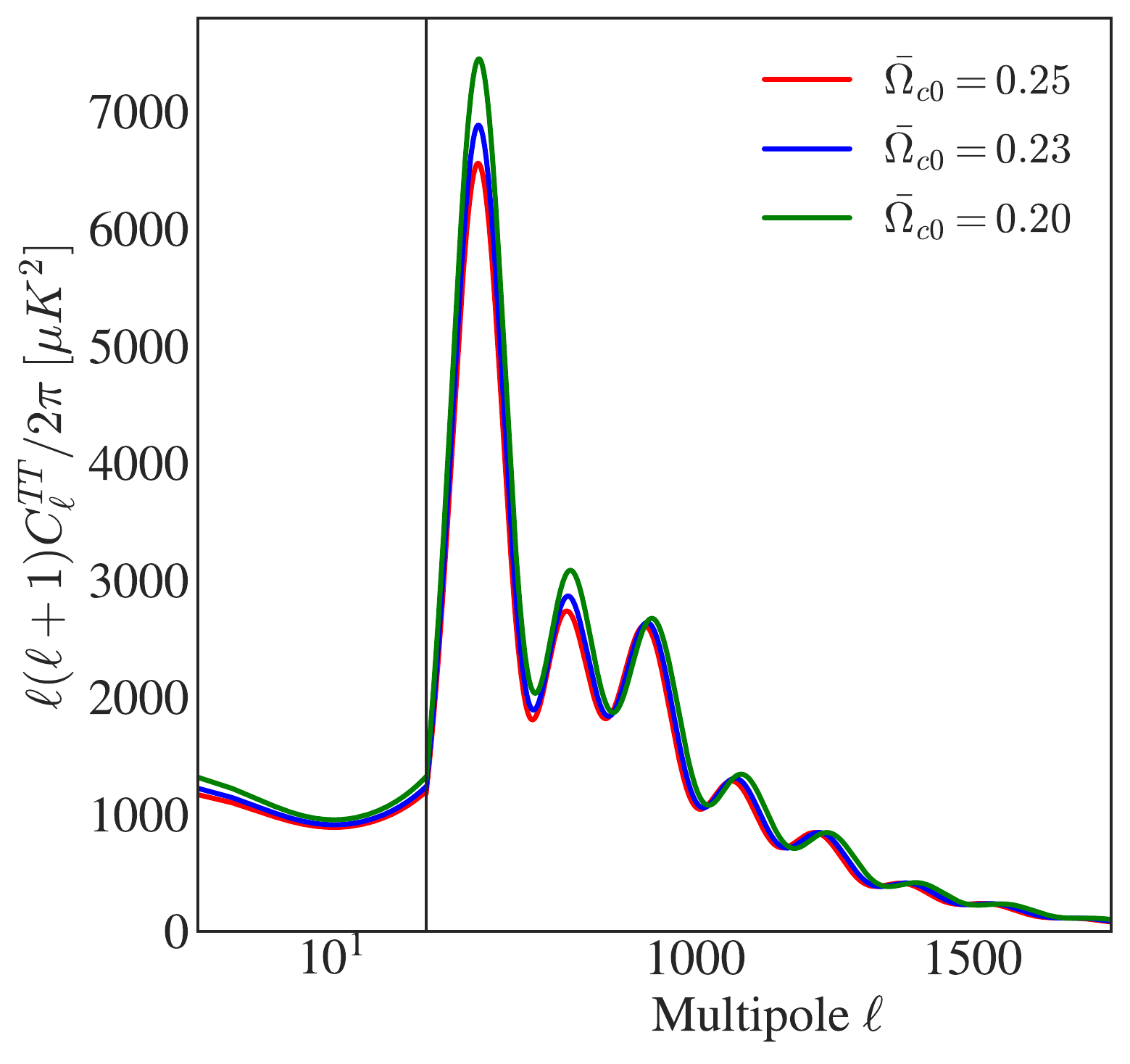}
\caption{CMB temperature anisotropies power spectrum for the $\bar{w}$gCg model. \textbf{Left panel: }Results obtained with different values of $w_0$, fixing $\tilde{\Omega}_{c0}=0.25$ (the value of $\bar{\Omega}_{c0}$ can be obtained from Eq.~\eqref{Omegabar}). \textbf{Right panel: }Results obtained with different values of $\bar{\Omega}_{c0}$, fixing $w_0=-0.95$.} 
\label{fig.cmb}
\end{figure}

To assess how perturbations grow in the $\bar{w}$gCg, we compute the temperature anisotropies in the CMB and the matter power spectrum for different values of the parameters $w_0$ and $\bar{\Omega}_{c0}$. In Fig.~\ref{fig.cmb} we show the CMB power spectrum for the $\bar{w}$gCg model. Whereas in the left panel we vary the parameter $w_0$, in the right panel we consider different values for $\bar{\Omega}_{c0}$. Regarding the left panel, we see that the model is very sensitive to $w_0$ values, offering changes in the format of the CMB power spectrum, mostly as the overall amplitude. In a general way, it is well-known that the overall amplitude of the CMB power spectrum depends on the time of the decoupling. Since the parameter $w_0$ affects the expansion dynamics of the Universe, i.e., it affects the Hubble rate, the strong sensitivity of the CMB power spectrum in relation to $w_0$ is expected. On the left panel, in addition to an increase in the amplitude, we can see the that the peak scales are slightly shifted to the left, which is compatible to a similar analysis in the context of the $\Lambda$CDM model.

\begin{figure}[h]
\centering
\includegraphics[width=0.49\columnwidth]{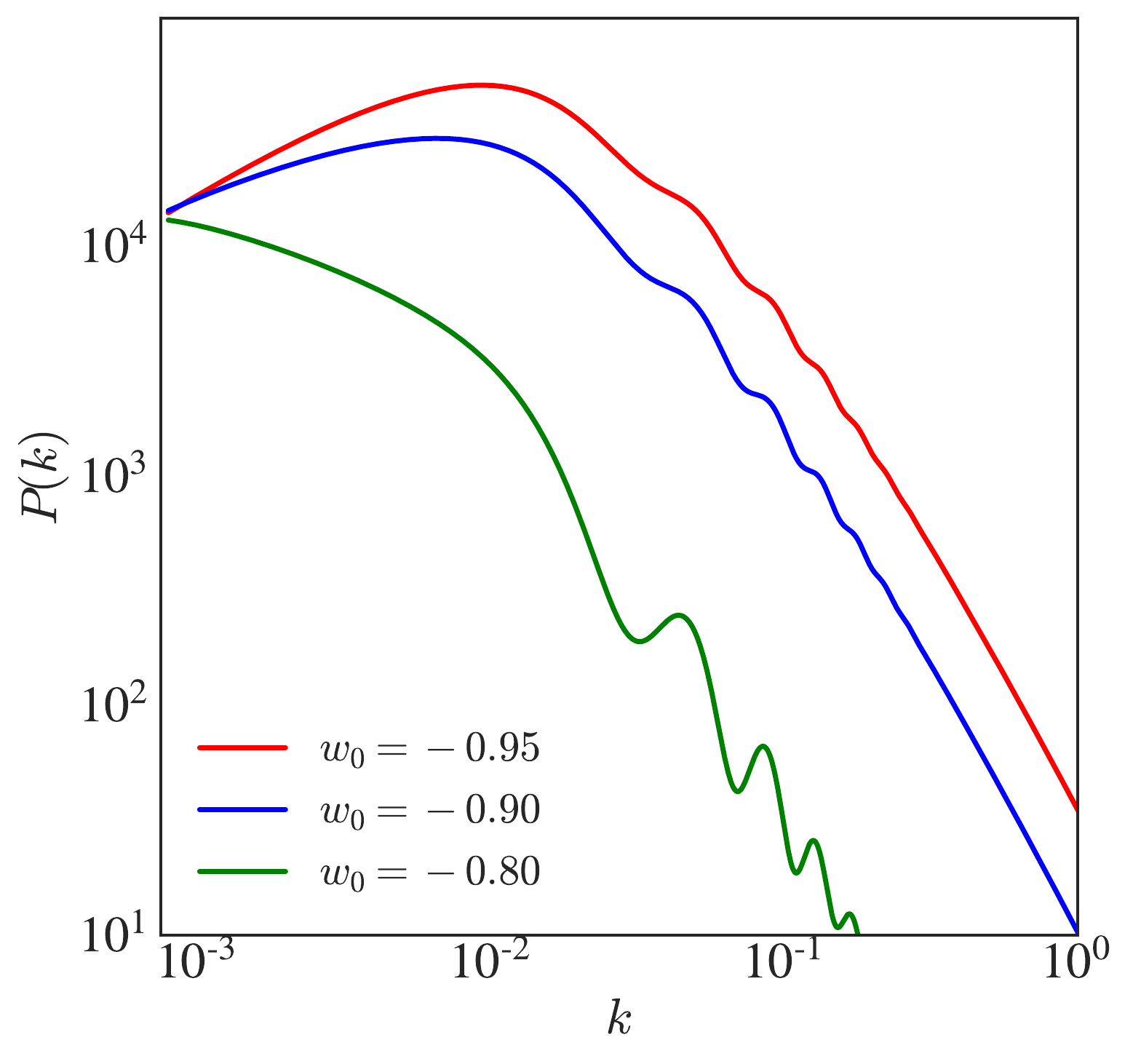}
\includegraphics[width=0.49\columnwidth]{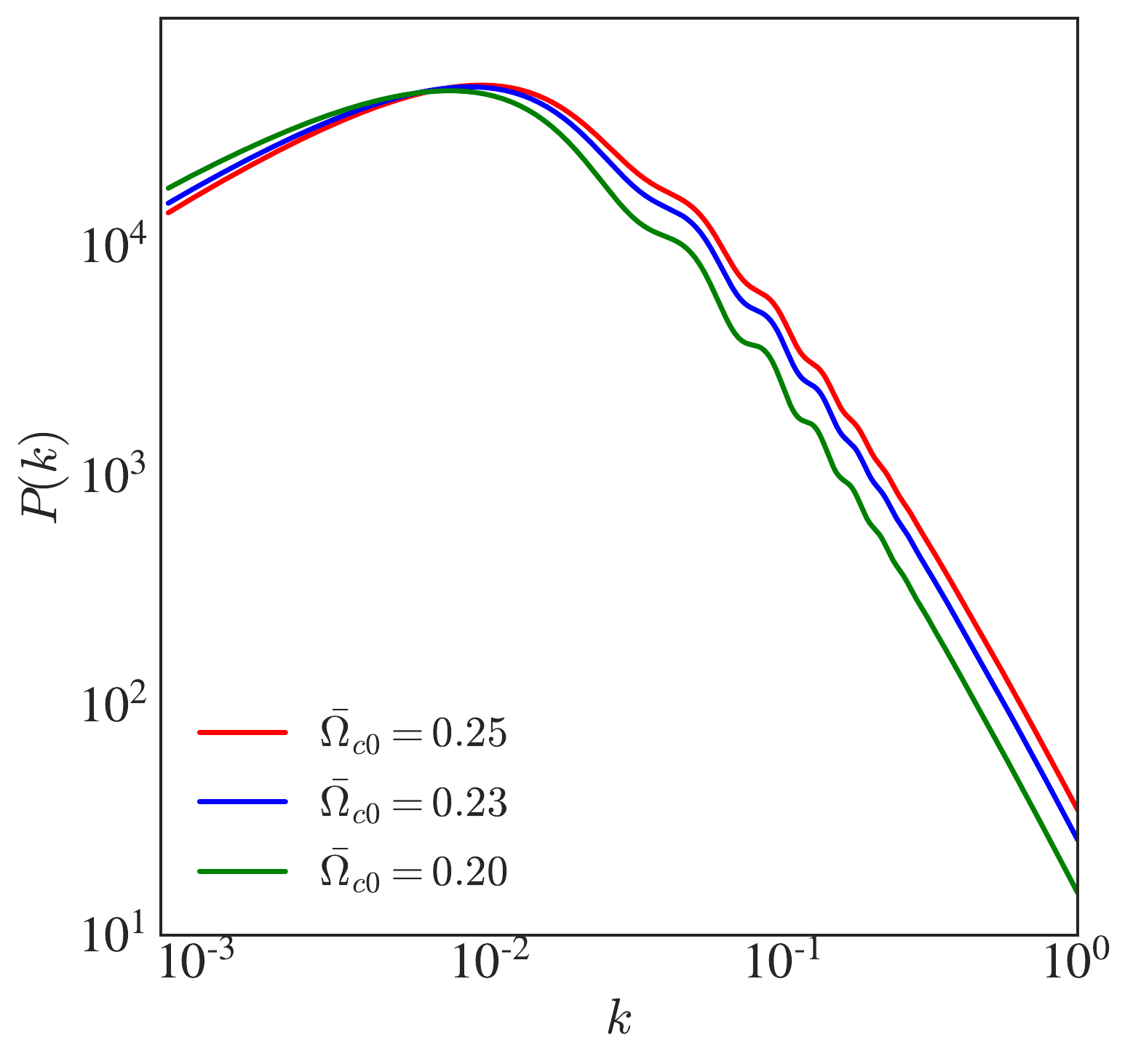}
\caption{Linear matter power spectrum for the $\bar{w}$gCg model. \textbf{Left panel: }Results obtained with different values of $w_0$, fixing $\tilde{\Omega}_{c0}=0.25$ (the value of $\bar{\Omega}_{c0}$ can be obtained from Eq.~\eqref{Omegabar}). \textbf{Right panel: }Results obtained with different values of $\bar{\Omega}_{c0}$, fixing $w_0=-0.95$.}
\label{fig.pk}
\end{figure}

Furthermore, we show in Fig.~\ref{fig.pk} the linear matter power spectrum for the $\bar{w}$gCg. Similarly to Fig.~\ref{fig.cmb}, the left panel shows different results obtained varying $w_0$, and the right panel shows different cases for $\bar{\Omega}_{c0}$. As well as the CMB spectrum, we can see from the left panel that the matter power spectrum is very sensitive to the $w_0$ parameter. One can conclude that, the bigger (less negative) the parameter $w_0$ is, the greater the suppression suffered by the matter power spectrum. To understand this behavior, we must remember that, as discussed in Sec.~\ref{ssec:dyncGc}, between $10^{0}<z<10^{3}$ the DE component behaves like pressureless matter. However, at the linear level, the sound speed of the DE component is luminal, while for the matter component the sound speed is zero. This fact implies that there is no DE clustering, even when it behaves like pressureless matter. Thus, the suppression of the power is due to the fact that only part of the cosmic substratum between $10^{0}<z<10^{3}$ is able to cluster. In light of this, we can conclude that our model works like if the parameter $w_0$ can prevent part of the matter in the Universe of clustering. Concerning the right panel, we can see the effect of varying $\bar{\Omega}_{c0}$ in the linear matter power spectrum. In this case, it is easy to see that the amplitude and the slope of the curve for large values of $k$ are slightly affected. It is well-known that these two effects are related to the matter density parameter. First, the amplitude\footnote{The amplitude of the linear power spectrum also depends on the amplitude of the primordial power spectrum, usually denoted by $A_s$.} is associated to the DE density parameter, which is given by $\bar{\Omega}_{x0}\approx 1-\bar{\Omega}_{c0}-\bar{\Omega_{b0}}$. On the other hand, it is also well-known that the slope depends on the ratio $\bar{\Omega_{b0}}/\bar{\Omega}_{c0}$.

\section{Statistical analysis}
\label{sec:stat_an}

\subsection{Cosmological data}

The next step after analyzing the $\bar{w}$gCg model from the theoretical perspective is to move to the observations. In this context, combined with the Boltzmann solver CLASS, we make use of the statistical code MontePython~\cite{Brinckmann:2018cvx} in order to perform a parameter selection with the recent data publicly available. In what follows we present the data used in this analysis:
\begin{itemize}
	\item \textbf{SNe Ia: }For the type Ia supernovae we employ the Pantheon sample~\cite{Scolnic:2018pan}, a compilation of 1048 apparent magnitude measurements, as well as its covariance matrix (including both systematic and statistical errors). The apparent magnitude, $m$, is related to the luminosity distance through the following relation:
\begin{equation}
	m=5\log_{10}\left[D_L\left(z\right)\right]+25+M\,,
\end{equation}
where $M$ is the absolute magnitude, and $D_L\left(z\right)$ is the luminosity distance. In the analysis, the absolute magnitude is considered as a nuisance parameter. The theoretical model enters in the luminosity distance.

	\item \textbf{BAO/RSD: }For the Baryon Acoustic Oscilation and Redshift Space Distortion, we consider the data from SDSS-DR7 Main Galaxy Sample~\cite{Ross:2014qpa}, BOSS-DR12 (LRG)~\cite{BOSS:2016wmc}, eBOSS-DR16 (LRG, QSO and Ly$\alpha$ auto- and cross-correlation with QSO)~\cite{eBOSS:2020yzd}. All these data have been obtained from galaxy/quasar clustering, however, instead of the full power spectrum, we consider here the compressed information contained in: ($i$) the angular BAO feature from $D_M/r_d$; ($ii$) the BAO radial feature from $D_H/r_d$; and the anisotropic features in galaxy clustering from $f\sigma_8$. In the BAO angular quantity, the characteristic distance is related to the angular diameter distance as $D_M\equiv\left(1+z\right)D_A$, and in the radial quantity, the characteristic distance is related to the Hubble rate via $D_H\equiv c/H$. Moreover, $r_d$ is the sound-horizon distance evaluated out to the baryon drag epoch, $f$ is the linear growth rate and $\sigma_8$ is the amplitude of mass fluctuations on scales on scale 8$h^{-1}$Mpc. In our analysis data from different catalogs are treated as independent, however we consider the covariance matrix for different measurements inside the same catalog. Overall, this covers a redshift range of $0.15<z<2.334$. A full table with all data used in this analysis can be found in~\cite{Andrade:2021njl}.
	
	\item \textbf{CMB: }For the CMB we use the Planck 2018 data with information from temperature, polarization and temperature polarization cross-correlation spectra (TT, EE, TE), as well as the lensing maps reconstruction~\cite{Aghanim:2020planck}. In this analysis we considered the standard likelihood codes: ($i$) the \texttt{COMANDER} likelihood version for low-$\ell$ TT spectrum, which contains data of the spectrum with $2\leq\ell<30$; ($ii$) the \texttt{SimAll} likelihood version for low-$\ell$ EE spectrum in the same interval of $\ell$; ($iii$) the Plick TTTEEE likelihood version for the TT spectrum with $30\leq\ell< 2500$ as well as the TE and EE spectra with $30\leq\ell<2000$; and ($iv$) the standard likelihood obtained from the lensing power spectrum reconstruction with $8\leq L\leq 400$. A detailed description of the likelihood codes can be found in Ref.~\cite{Planck:2019nip}.
	 	
	\item \textbf{Weak Lensing: }For the weak lensing, we use the data from the KiDS-1000~\cite{Kuijken:2019gsa,Giblin:2020quj}. In this case, the relevant cosmological observable is the weak lensing power spectrum $\xi_{\pm}\left(\theta\right)$ for the auto- and cross-correlations across four tomographic redshift bins. Following the KiDS-1000 orientation, we make use of the COSEBIs (Complete Orthogonal Sets of E/B-Integrals~\cite{Schneider:2001af}) as our summary statistic. 
	
\end{itemize}

\subsection{Results}
\label{ssec:results}

\begin{figure}[h]
\centering
\includegraphics[width=0.7\textwidth]{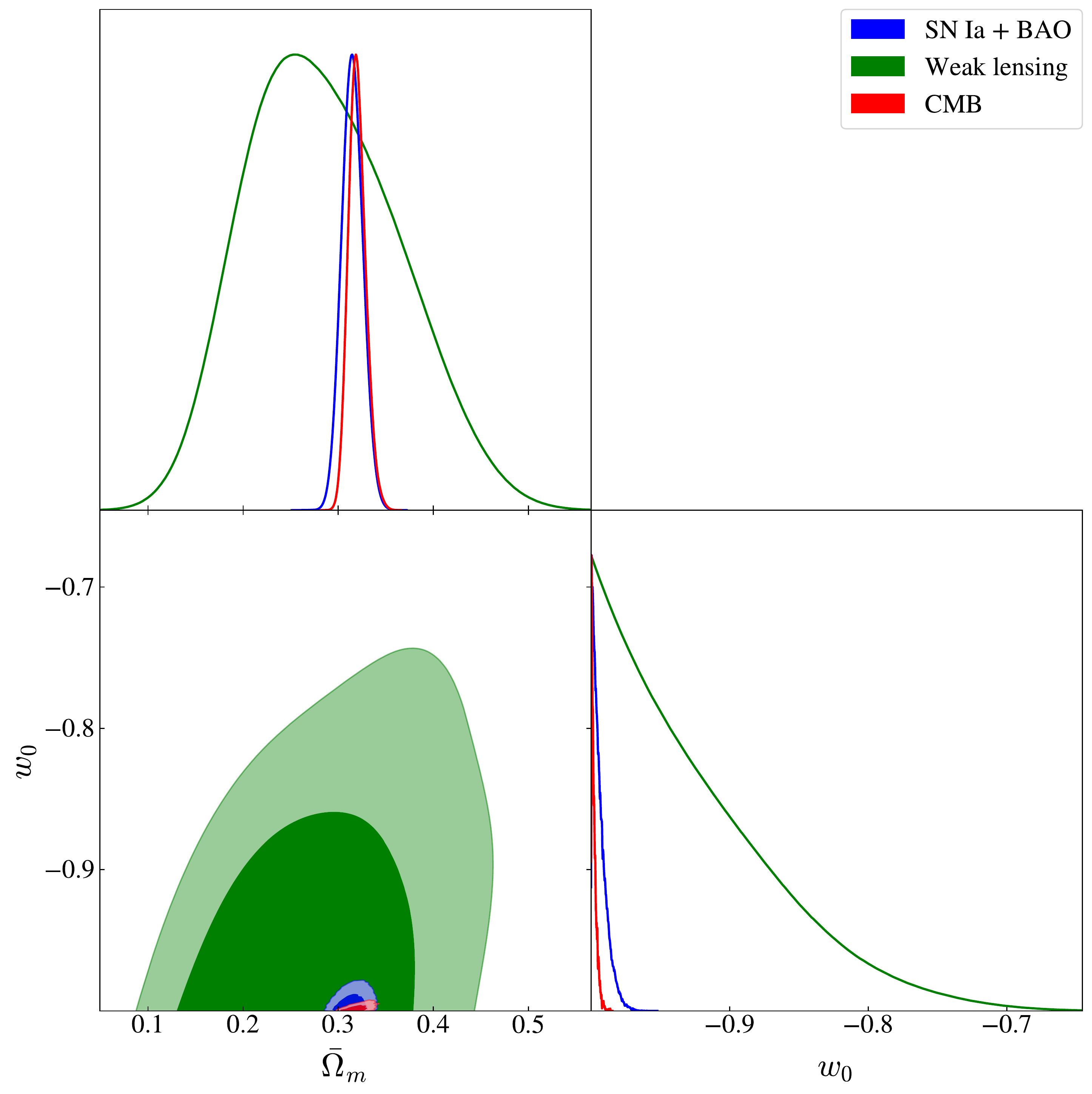}
\caption{Corner plot of $\bar{\Omega}_m-w_{0}$ for the statistical analysis for the $\bar{w}$gCg model. Each dataset is used in a different color: Weak lensing (green), SN Ia + BAO (blue) and CMB (red). The inner and outer regions represent $1\sigma$ and $2\sigma$ CL, respectively. We only find lower and upper limits for values of $w_{0}$, for which we find that the best constrains are offered by CMB and SN Ia + BAO, whereas weak lensing provides weak constraining power. }
\label{fig.triangle}
\end{figure}

In this section we present the results of our statistical analysis for the $\bar{w}$gCg model. We  combine the contributions for SN Ia and BAO, as the background data (in blue), and maintain the CMB and weak lensing separated in order to study the impact that each dataset has on the model. In Fig.~\ref{fig.triangle} we show the corner plot for the parameters $\bar{\Omega}_m$ and $w_0$. As can be seen, the SN Ia+BAO and CMB have the most constraining power for most of variables of interest. %This fact is not a surprise, since we have seen in Figs.~\ref{fig.wx} and~\ref{fig.cmb} that our model is indeed very sensitive to this parameters (specially $w_0$). %On the other hand, the weak lensing data provides a weak constraint. This is not a surprise also, since it is well-known that weak lensing data can not constraint strongly parameters related with a dynamical evolution of the DE component.     

The results of all analysis performed are presented in Tab.~\ref{tab.results}. %For the sake of comparison, we also present the results obtained for the $w$CDM model, since both cases represent a one-parameter description for the DE component. 
From Table~\ref{tab.results} we can observe that aside from weak lensing, we find that $w_0$ is tightly constrained. Note that, since $\bar{w}\left(z\right)\geq-1$ for all values of $z$, our parameter selection only provides a upper-bound for $w_0$, being -1 its lower-bound. As we can see in Tab.~\ref{tab.results}, the constrains provided by the data from SN Ia+BAO and CMB allows deviations in relation to $\Lambda$CDM less them 1\%. Such  strong constraints indicate that our model affects considerably the CDM component at early times, which affects both the CMB and BAO physics. The joint analysis is not shown because it is fully dominated by the CMB data. It is important to mention that our study is in agreement with Ref.~\cite{Supriya:2020int}, where a study for different cases of interacting models, including the gCg, is performed. Finally, it worth emphasizing that small deviations from $\Lambda$CDM does not necessarily exclude the model here explored~\cite{Castello:2022cau}.
\begin{table*}[]
\renewcommand{\arraystretch}{1.3}
\centering
\begin{tabular}{c c c c c}
    \hline \hline
    \multicolumn{5}{c}{$\bar{w}$gCg} \\ \hline    
    %\hline
     & $w_0$ & $\bar{\Omega}_m$ & $H_0$ & $S_8$ \\
    \hline
     SN Ia + BAO & $< -0.992$ & $0.315 \pm 0.011$ & $70.5^{+1.4}_{-1.8}$  & - \\
     Weak Lensing & $< -0.902$ & $0.282^{+0.071}_{-0.090}$ & $72.9 \pm 5.0$ & $0.828^{+0.042}_{-0.073}$\\
     CMB & $< -0.997$ & $0.320^{+0.008}_{-0.010}$ & $66.90^{+0.72}_{-0.60}$ & $0.833 \pm 0.013$\\ [1ex]
%    \hline
%    \multicolumn{5}{c}{$\bar{w}$CDM} \\
%     & $w_0$ & $\bar{\Omega}_m$ & $H_0$ & $S_8$ \\
%    \hline
%     SN Ia + BAO & $-1.01_{-0.009}^{+0.008}$ & $0.258_{-0.0157}^{+0.014}$ & $68.5 \pm 0.25$  & - \\
%     Weak Lensing &  &  &  & \\
%     CMB &  &  &  & \\
    \hline
    \hline
\end{tabular}
\caption{Best fit and $1\sigma$ confidence level obtained for each sample used. Current constrains for our model only offer upper limits for $w_0$, which is also evident in Fig.~\ref{fig.triangle}. Aside from weak lensing we find tigh constrains for the other parameters. We also find that the CMB data shows preference for higher values of $H_0$ when in comparison to $\Lambda$CDM scenario.}
\label{tab.results}
\end{table*}

In order to assess how the model responds to the current observational tensions of the standard cosmology, we also compare the values obtained for $H_0$ with the $H_0$ value obtained by~\cite{Riess:2020fzl}, $H_0 = 73.2\pm 1.3$ km/s/Mpc. In what concerns the joint analysis with CMB, we find that the tension remains almost identical, due to the weight of the CMB data on the estimate of $H_0$. On the other hand, regarding the $S_8$ tension, when we compare the results for the plane $\Omega_m-S_8$ from Planck 2018 and KiDS-1000, we find an interesting result. In this case, our model delivers predictions compatible with higher values of $S_8$ from weak lensing. This result makes both analyses from CMB and weak lensing compatible within $\simeq 1\sigma$ level.
\begin{figure}[h]
\centering
\includegraphics[width=0.49\textwidth]{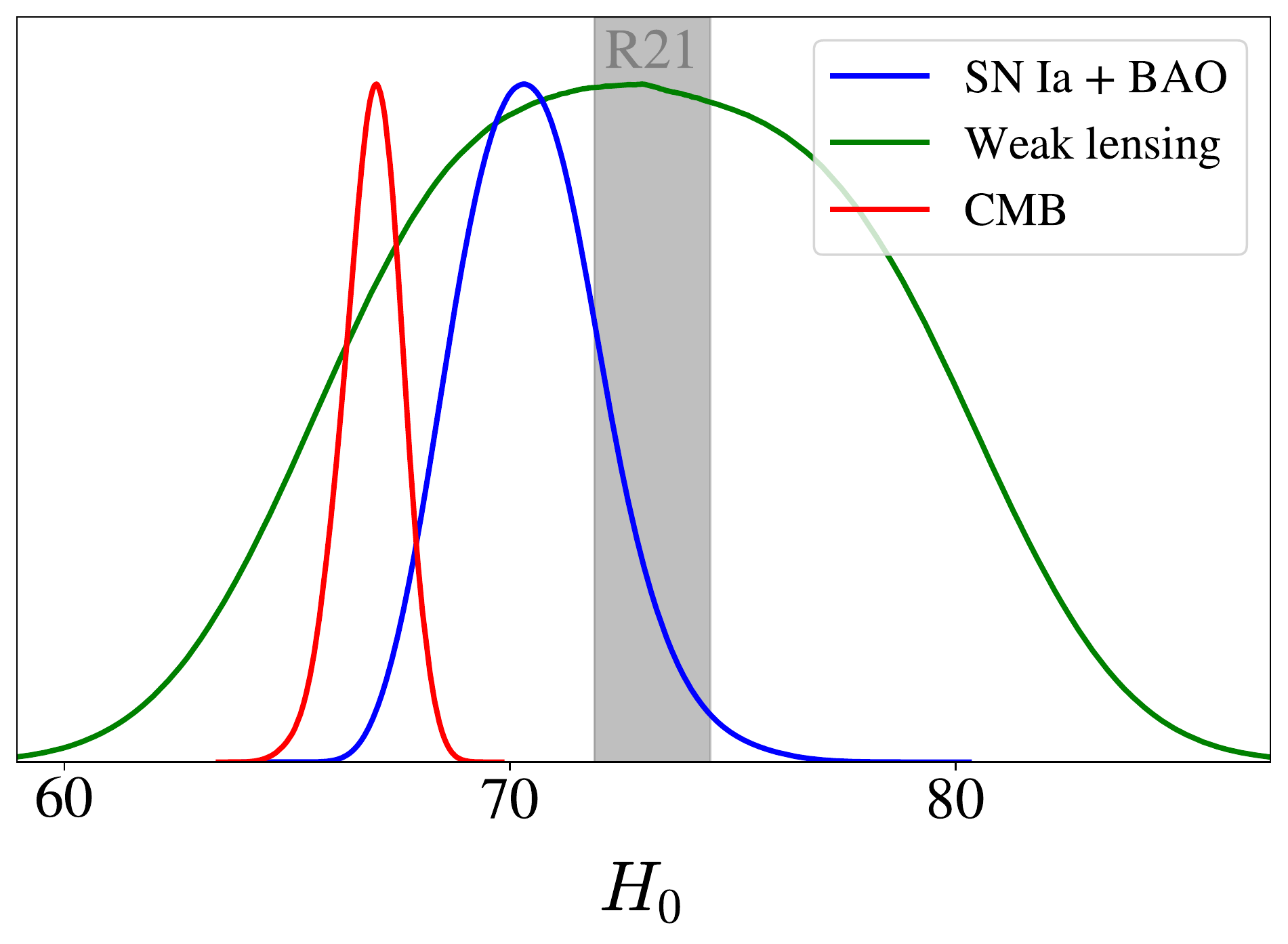}
\includegraphics[width=0.49\textwidth]{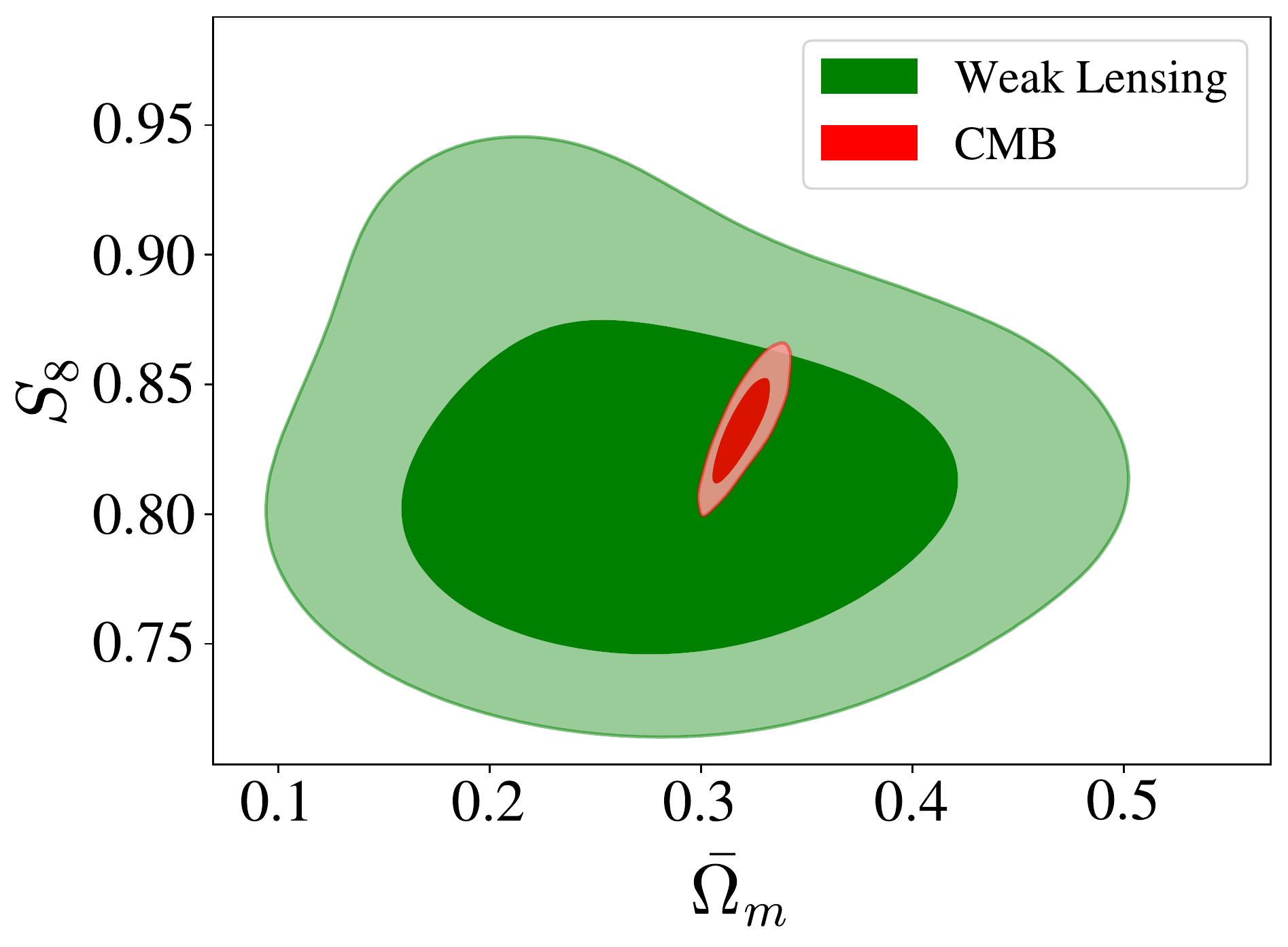}
\caption{$H_{0}$ and $S_{8}$ distributions for $\bar{w}$CDM model. \textbf{Left panel:} $H_{0}$ posteriors for each dataset is presented in a different color: SN Ia + BAO (blue), weak lensing (green), CMB (red) for $1\sigma$. The grey region represents the bounds for the value found in Ref.~\cite{Riess:2020fzl}. As seen in Tab.~\ref{tab.results}, the CMB possesses the highest constraining power for $H_{0}$, followed by SN Ia + BAO, while weak lensing provides the worst constraining power for the same parameter. %In regards to R21, we find an improvement as SN Ia + BAO data are in agreement for $1\sigma$, and although this is not the case for CMB data, we recall that the tension in $H_{0}$ is alleviated, as seen in Tab.~\ref{tab.results}. 
\textbf{Right panel:} $S_8-\bar{\Omega}_{m}$ plane obtained from weak lensing and CMB analysis. Once again, the inner and outer regions represent $1\sigma$ and $2\sigma$ CL. We call attention to the fact that both regions are in agreement for $1\sigma$, reducing the $S_8$ tension from $3\sigma$ to $1\sigma$ CL.}
\label{fig.H0}
\end{figure}

\section{Conclusions} 
\label{sec:conclusions}

In the past decade, several authors have debated possibilities to describe the dynamics of the current phase of the accelerated expansion of the Universe. Here we have followed Ref.~\cite{vonMarttens:2019ixw} and studied how an interaction based on the generalized Chaplygin gas could act as a dynamical DE model. We showed that the dark degeneracy mapping delivers a time varying one-parameter EoS parameter for the DE component. Particularly, this result has two remarkable features: ($i$) we found a dynamical evolution for the DE component with only one parameter; and ($ii$) the resulting $\bar{w}_x\left(z\right)$ naturally does not cross the phantom line, thereby satisfying $0>\bar{w}_x\left(z\right)\geq\bar{w}_x\left(z=0\right)=w_0$ and avoiding instabilities. Another important characteristic of the model is that at early times the DE EoS parameter tends to zero, i.e., the DE component behaves like a pressureless component. Considering a DE luminal sound speed, it affects the clustering in the matter-dominated epoch. In that sense, the model works as a mechanism that prevents all pressureless component of clustering. 

From the observational perspective, we showed that the current available data sets, specially CMB, provide strong constraints on the extra parameter $w_0$, in good agreement with the $\Lambda$CDM model. We also briefly discussed the current tensions of cosmology in light of the model, and found that while it does not have much impact on the $H_0$ values, it favors higher values of $S_8$, making the CMB and weak lensing data analysis compatible within $\simeq 1\sigma$. % levelalthough confidence levels are worse when compared to the standard model.

%So far, the current cosmological data is not conclusive about a preferred alternative model. However, recent forecasts such as Ref.~\cite{Salzano:2021j,Figueruelo:2021elm} made by J-PAS search team, show promising results for constraining dynamical models. In general, upcoming experiments such as J-PAS, DESI and Euclid~\cite{Blanchard:2020euclid,Euclid:2022euclid} will provide us with higher quality data which can help us settle the question and may shed some much needed light for unraveling the dark energy nature.  

%-------------------------------------------
\acknowledgments
RvM is supported by the Programa de Capacita\c{c}\~ao Institucional PCI/ON. DB acknowledges financial support from the Coordena\c{c}\~ao de Aperfei\c{c}oamento de Pessoal de N\'{\i}vel Superior - CAPES. JSA is supported by Conselho Nacional de Desenvolvimento Cient\'{\i}fico e Tecnol\'ogico (CNPq 310790/2014-0) and FAPERJ grant 259610 (2021). 
%-------------------------------------------

\bibliographystyle{JHEP}
\bibliography{mybib}

\appendix

\section{Dynamical DE parameterization}
\label{ssec:dynDE}

The usual approach for modeling DE scenarios consists on the hypothesis that the components of the dark sector are independent and the DE component has a dynamical (time-dependent) EoS parameter $w_{x}\left(a\right)$. In this approach, a number of DE parameterizations have been proposed and studied in the current literature. In order to distinguish the dynamical approach from models with interaction, we follow the convention adopted in Ref.~\cite{vonMarttens:2019ixw} and denotes all quantities with a bar.

Since the matter species are independent, the background energy conservation is given by,
\begin{eqnarray}
\dot{\bar{\rho}}_{c}+3\bar{H}\bar{\rho}_{c}&=&0 \,, \label{drhoc1} \\
\dot{\bar{\rho}}_{x}+3\bar{H}\bar{\rho}_{x}\left[1+\bar{w}_{x}\left(a\right)\right]&=&0 \,. \label{drhox1}
\end{eqnarray}
These expressions have the well-known solutions,
\begin{eqnarray}
\bar{\rho}_{c}&=&\frac{8\pi G}{3H_{0}}\bar{\Omega}_{c0}a^{-3} \,, \label{rhoc1} \\
\bar{\rho}_{x}&=&\frac{8\pi G}{3H_{0}}\bar{\Omega}_{x0}\exp\left[-3\int\dfrac{1+\bar{w}_{x}\left(a^{\prime}\right)}{a^{\prime}}da^{\prime}\right] \,. \label{rhox1}
\end{eqnarray}
Thus, using the results from Eqs.~\eqref{rhoc1} and~\eqref{rhox1}, the ratio between CDM and DE energy densities is given by
\begin{equation}
\bar{r}\left(a\right)=\bar{r}_{0}a^{-3}\exp\left[3\int\dfrac{1+\bar{w}_{x}\left(a^{\prime}\right)}{a^{\prime}}d\bar{a}\right]\,,
\end{equation}
where $\bar{r}_{0}=\bar{\rho}_{c0}/\bar{\rho}_{x0}=\bar{\Omega}_{c0}/\bar{\Omega}_{x0}$ is the current value of the ratio $\bar{r}$.

Lastly, the unified dark EoS parameter is given by,
\begin{equation}\label{wd1}
\bar{w}_{d}\left(a\right)=\dfrac{\bar{w}_{x}\left(a\right)}{1+\bar{r}_{0}a^{-3}\exp\left[3\int\dfrac{1+\bar{w}_{x}\left(a^{\prime}\right)}{a^{\prime}}da^{\prime}\right]} \,.
\end{equation}
The Eq.~\eqref{wd1} reflects the fact that, in this case, where no dark interaction is taken into account, it is straightforward to conclude that $w_{x}\left(a\right)$ alone defines the background dynamics of the dark sector uniquely.

\section{Interacting dark sector}
\label{ssec:intDE}

We shall now relax the hypothesis that dark components are independent, and we assume a phenomenological source term responsible for energy transfer in the background. Again, following the convention established in the Ref.~\cite{vonMarttens:2019ixw}, quantities related to the interacting approach will be denoted with a tilde. In contrast with the fact that dark components are allowed to interact with each other, we now fix the DE EoS parameter as $\tilde{w}_{x}=-1$. In this context, the energy balance of the dark components are given by
\begin{eqnarray}
\dot{\tilde{\rho}}_{c}+3\tilde{H}\tilde{\rho}_{c}&=&\tilde{Q}\,, \label{cdmenergy} \\
\dot{\tilde{\rho}}_{x}&=&-\tilde{Q} \,, \label{deenergy}
\end{eqnarray}
where $\tilde{Q}$ is a source function that describes an energy exchange between the dark components. Whereas the magnitude of $\tilde{Q}$ is related to the strength of the interaction, its sign defines the direction of the energy transfer. If $\tilde{Q}$ is positive one finds a decaying DE component leading to matter creation, and if $\tilde{Q}$ is negative the opposite occurs.

As previously mentioned, the ratio between CDM and DE energy densities is associated with interaction in the dark sector. To make it clear, it is convenient to introduce the ratio and its derivative with respect to cosmic time, 
\begin{equation} \label{drint}
\tilde{r}\equiv\dfrac{\tilde{\rho}_{c}}{\tilde{\rho}_{x}}\quad\Rightarrow\quad\dot{\tilde{r}}=\tilde{r}\left[\tilde{Q}\left(\dfrac{\tilde{\rho}_{c}+\tilde{\rho}_{x}}{\tilde{\rho}_{c}\, \tilde{\rho}_{x}}\right)-3\tilde{H}\right]\,.
\end{equation}
The Eq.~\eqref{drint} is a differential equation for $\tilde{r}$ in terms of the interaction term $\tilde{Q}$, which, in general, is phenomenologically determined.

As an \textit{ansatz}, we assume that the interaction function is written as $\tilde{Q}=3\tilde{H}\tilde{\gamma}\tilde{R}\left(\tilde{\rho}_{c},\tilde{\rho}_{x}\right)$, where $\tilde{H}$ is the Hubble rate, $\tilde{\gamma}$ is the interacting parameter and $\tilde{R}$ is a general function of the dark components' energy densities. Thus, the equation \eqref{dr} can be rewritten as,
\begin{equation} \label{drtilde}
\dot{\tilde{r}}-3\tilde{H}\tilde{r}\left[\tilde{\gamma}\tilde{R}\left(\dfrac{\tilde{\rho}_{c}+\tilde{\rho}_{x}}{\tilde{\rho}_{c}\, \tilde{\rho}_{x}}\right)-1\right]=0\,.
\end{equation}
Assuming that there is no privileged energy density scale in the Universe, the first term into the square brackets must be a function of $\tilde{r}$,
\begin{equation} \label{frint}
\tilde{f}\left(\tilde{r}\right)\equiv \tilde{\gamma}\tilde{R}\left(\frac{\tilde{\rho}_{c}+\tilde{\rho}_{x}}{\tilde{\rho}_{c}\, \tilde{\rho}_{x}}\right)\,,
\end{equation}
where, in a FLRW spacetime, it must be a function of the cosmic time, or equivalently, a function of the scale factor. Using the definition \eqref{frint}, Eq.~\eqref{drtilde} can be rewritten as,
\begin{equation} \label{dr}
\dot{\tilde{r}}-3\tilde{H}\tilde{r}\big[\tilde{f}\left(\tilde{r}\right)-1\big]=0 \,.
\end{equation}
An specific interacting model is, in general, proposed by an phenomenological choice of one of the functions: $\tilde{Q}$, $\tilde{r}\left(a\right)$ or $\tilde{f}\left(\tilde{r}\right)$. The Eqs.~\eqref{drint},~\eqref{drtilde},~\eqref{frint} and~\eqref{dr} exhibit how these functions are not independent of each other, but they define only one degree of freedom. For example, an interacting model described by the interaction function $\tilde{Q}=3\tilde{H}\tilde{\gamma}\tilde{\rho}_{c}^{\alpha}\tilde{\rho}_{x}^{\beta}\left(\tilde{\rho}_{c}+\tilde{\rho}_{x}\right)^{\sigma}$ can be related to the following $\tilde{f}\left(\tilde{r}\right)$ function,
\begin{equation}
f\left(\tilde{r}\right)=\tilde{r}^{\alpha-1}\left(1+\tilde{r}\right)^{\sigma+1}\,.
\end{equation}

Finally, according Eq.~\eqref{EoS}, the unified dark EoS parameter for a given interacting model is given by,
\begin{equation} \label{wd2}
\tilde{w}_{d}=-\dfrac{1}{1+\tilde{r}\left(a\right)}\,.
\end{equation}

\end{document}